\documentclass[usenatbib]{mn2e}

\usepackage{lscape,times,graphicx}

\newcommand{\microns}{$\mu$m}

\def\Pa{P\,$\alpha$}
\def\hei{He\,{\sc i}}
\def\heii{He\,{\sc ii}}
\def\civ{C\,{\sc iv}}
\def\niii{N\,{\sc iii}}
\def\niv{N\,{\sc iv}}

\def\oiv{O\,{\sc iv}}

\def\ga{\mathrel{\hbox{\rlap{\hbox{\lower4pt\hbox{$\sim$}}}\hbox{$>$}}}}
\def\la{\mathrel{\hbox{\rlap{\hbox{\lower4pt\hbox{$\sim$}}}\hbox{$<$}}}}
\def\msunyr{$M$ \mbox{$_{\normalsize\odot}$}\rm{yr}$^{-1}$}
\def\msun{M\mbox{$_{\normalsize\odot}$}}

\def\lsun{L\mbox{$_{\normalsize\odot}$}}
\def\kms{\,km~s$^{-1}$}

\def\arcsec{$^{\prime \prime}$}
\def\arcmin{$^{\prime}$}

\def\hii{H{\sc ii}}

\def\lsun{$L$\mbox{$_{\normalsize\odot}$}}
\def\lstar{$L$\mbox{$_{\star}$}}
\def\Rstar{\hbox{$R_*$}}
\def\Mdot{\hbox{$\dot {M}$}}
\def\Lstar{\hbox{$L_*$}}
\def\Vinf{\hbox{$v_\infty$}}
\def\kms{\,km~s$^{-1}$}

\def\arcsec{$^{\prime \prime}$}
\def\arcmin{$^{\prime}$}
\def\mgii{Mg\,{\sc ii}}
\def\hi{H\,{\sc i}}
\def\um{$\mu$m}

\def\MSX{{\it MSX}}
\def\Pa{P$\alpha$}
\def\vlsr{$v_{\rm lsr}$}

\newcommand{\fig}[1]{Fig.\ \ref{#1}}
\newcommand{\Fig}[1]{Figure \ref{#1}}

\title[A young massive star cluster at the end of the Galactic Bar]{A newly-discovered
  young massive star cluster at the far end of the Galactic Bar}  
\author[B. Davies et al.]
{Ben~Davies$^{1,2}$,
  Diego~de~la~Fuente$^{3}$,
  Francisco~Najarro$^{3}$, 
  Jim~A.~Hinton$^{4}$\newauthor
  Christine Trombley$^{5}$,
  Donald~F.~Figer$^{5}$,
  Elena~Puga$^{3,6}$ \\
$^{1}$Institute of Astronomy, University of Cambridge,
  Madingley Road, Cambridge CB3 0HA, UK.\\
$^{2}$School of Physics \& Astronomy, University of Leeds,
  Woodhouse Lane, Leeds LS2 9JT, UK.\\
$^{3}$Centro de Astrobiologia (INTA/CSIC), Instituto Nacional
  de Tecnica Aeroespacial, Ctra. de Ajalvir, km.~4, 28850 Torrejon de
  Ardoz, Madrid, Spain.\\
$^{4}$Department of Physics and Astronomy, University of
  Leicester, Leicester, LE1 7RH, UK.\\
$^{5}$Center for Detectors, Rochester
Institute of Technology, 54 Lomb Memorial Drive, Rochester NY, 14623,
USA. \\
$^{6}$Herschel Science Centre, European Space Astronomy Centre,
P.O.~Box~78, 28691~Villanueva de la Ca\~{n}ada, Madrid, Spain 
}
\begin{document}

\date{Accepted ... Received ...}

\pagerange{\pageref{firstpage}--\pageref{lastpage}} \pubyear{2009}

\maketitle

\label{firstpage}

\begin{abstract}
We present a near-infrared study of the candidate star cluster
Mercer~81, located at the centre of the G338.4+0.1 \hii\ region, and
close to the TeV gamma-ray source HESS~1640-465. Using HST/NICMOS
imaging and VLT/ISAAC spectroscopy we have detected a compact and
highly extincted cluster of stars, though the bright stars in the
centre of the field are in fact foreground objects. The cluster
contains nine stars with strong \Pa\ emission, one of which we
identify as a Wolf-Rayet (WR) star, as well as an A-type
supergiant. The line-of-sight extinction is very large, $A_{V}\sim
45$, illustrating the challenges of locating young star clusters in
the Galactic Plane. From a quantitative analysis of the WR star we
argue for a cluster age of 3.7$^{+0.4}_{-0.5}$\,Myr, and, assuming
that all emission-line stars are WRs, a cluster mass of $\ga
10^4$\msun.  A kinematic analysis of the cluster's surrounding
\hii-region shows that the cluster is located in the Galactic disk at
a distance of 11$\pm$2\,kpc. This places the cluster close to where
the far end of the Bar intersects the Norma spiral arm. This cluster,
as well as the nearby cluster [DBS2003]179, represent the first
detections of active star cluster formation at this side of the Bar,
in contrast to the near side which is well known to have recently
undergone a $\sim 10^6$\msun\ starburst episode.

\end{abstract}



\section{Introduction} \label{sec:intro}
Young massive star clusters (YMCs -- ages $\la$ 50Myr, masses $\ga
10^4$\msun) have relevance to many areas of astrophysics. They contain
large numbers of massive stars, whose high temperatures, high
luminosities, dense winds, and supernova explosions make YMCs a
considerable source of mechanical energy, ionizing radiation and
chemically processed ejecta. The effect that they have on their
surroundings is profound, clearing away the remains of their natal
molecular cloud, whilst revealing and triggering subsequent
generations of star formation
\citep[e.g.][]{Gonzalez-Delgado00,Smith06_census,Danks-paper,DeMarchi11}. Their
large populations of massive stars make them ideal natural
laboratories with which to study the evolution of massive stars up to
supernova and beyond
\citep[e.g.][]{Martins07,Martins08,Bibby08,RSGC1paper,SGR1900paper}. Finally,
they can dominate the radiative output of their host galaxies, either
through direct ultraviolet and optical emission, or through
reprocessed emission in the form of ionised gas or heated dust
\citep{Alonso-Herrero02}.

Our knowledge of our own Galaxy's population of YMCs is extremely
incomplete, in stark contrast to external galaxies
\citep[e.g.][]{Bastian05,Konstantopoulos09}. The high levels of
interstellar extinction in the plane of the Galaxy have meant that
until recently very few clusters were known beyond a distance of
$\sim$1kpc. Most known clusters beyond this distance were found by
targeted searches of, for example, the Galactic Centre
\citep{Cotera96,Figer99_GC}, giant \hii-regions
\citep{Blum99,Blum00,Blum01}, fields around newly-born neutron stars
\citep{Fuchs99,Vrba00}, or simply because the foreground extinction
was low enough to detect the cluster at optical wavelengths
\citep{Westerlund87}. However, recent infrared (IR) surveys of the Galactic
plane, beginning with 2MASS \citep{Skrutskie06}, and more recently
Spitzer/GLIMPSE \citep{Benjamin03} and VVV \citep{VVV}, are at last
helping us to uncover the cluster population of the Galactic disk, and
affording the opportunity to search the Galactic Plane for clusters in
a systematic way.

By-eye and algorithmic searches of survey data
\citep[e.g.][]{Ivanov02,Dutra03,Mercer05,Froebrich07,Borissova11} have
yielded over 1,000 candidates for newly discovered star
clusters. However, such catalogues inevitably contain large numbers of
false positives, due to chance alignments of stars, patchy foreground
extinction, and spatially extended emission incorrectly classified as
unresolved star clusters. Therefore, these catalogues of candidates
must be analysed carefully using multiwavelength survey data and,
ultimately, follow-up spectroscopic observations before their
distances and physical properties may be derived
\citep[e.g.][]{Kurtev07,Messineo09,Hanson10}. Only then can they be
placed in the framework of the Galaxy's recent star-forming history.

\begin{figure*}
  \centering
  \includegraphics[width=18cm,bb=60 15 955 449,clip]{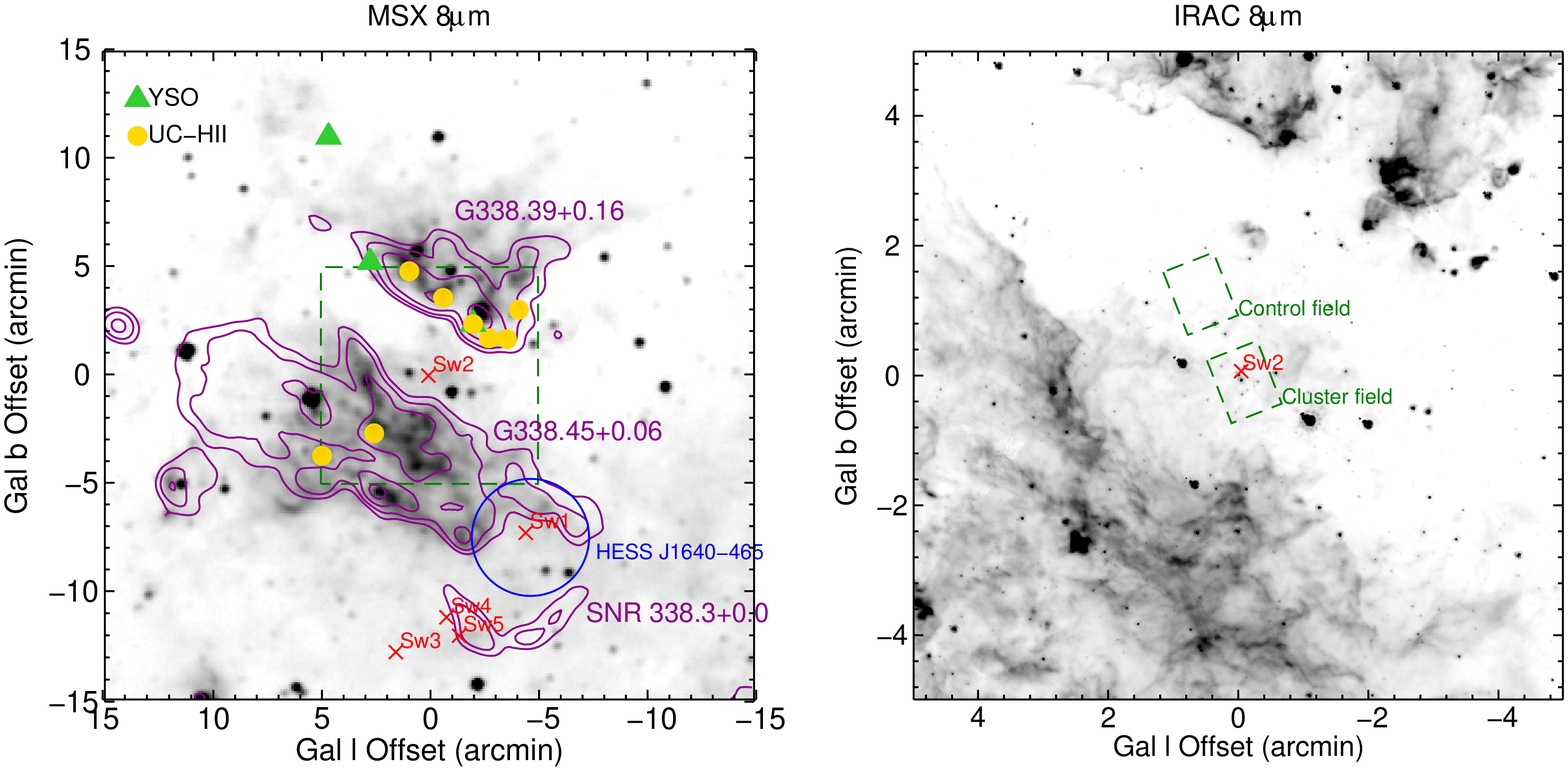}
  \caption{Wide-field images of Mc81 and its surroundings. {\it Left}:
    the MSX 8\um\ image, which shows the diffuse nebula surrounding
    the cluster. The locations of {\it Swift} X-ray sources are
    indicated \citep{Landi06}, as well as the location of the TeV
    emission source HESS J1640-465. The contours indicate the
    morphology of the 843MHz emission \citep{SUMSS}, and the dashed
    green box shows the field-of-view of the right-hand panel. {\it
      Right}: a higher resolution 8\um\ image from the GLIMPSE
    survey. The brightest stars of the cluster are coincident with the
    X-ray source Sw~2. The NICMOS cluster and control fields of view
    are illustrated by the dashed green boxes.}
  \label{fig:wfim}
\end{figure*}

One such candidate is object \#81 from the catalogue of
\citet{Mercer05}, known hereafter as Mc81\footnote{Objects in this
  catalogue are referred to by some authors with the prefix GLIMPSE,
  e.g.\ {\it GLIMPSE81}}. This object was found in an algorithmic
search of the GLIMPSE survey, by looking for spatial groups of stars
with similar photometric properties. The clues to its nature, however,
come from cross-correlation with other data in the literature. The
object appears to be at the centre of the \hii-region G338.4+0.1, a
bubble of warm dust and ionized gas, visible in the SUMSS 843MHz
\citep{SUMSS} and \MSX\ 8\um\ \citep{MSX} images, and in more detail
in the GLIMPSE 8\um\ image (\fig{fig:wfim}). The object's location is
close to a supernova remnant (SNR) by \citet{Green04} from the `shell'
morphology of continuum radio emission, but with no spectral index
measurement the object could also be a wind-blown bubble.

Close to the centre of the SNR is a high-energy TeV gamma-ray source,
HESS~J1640-465 \citep{Aharonian05}, at a distance of
8\arcmin\ (a linear distance of 22pc, if the complex is at a distance
of 11kpc -- see Sect.\ \ref{sec:extinct}). This source, as with many other TeV
sources, is thought to be associated with a pulsar wind nebula
\citep{Funk07}, indicative of recent SN activity. Indeed, TeV emission
can be a useful tracer of massive star formation in the Galactic Plane
as it is unaffected by absorption, and there are other young massive
star clusters in the literature that are known to be associated with
such sources -- RSGC1 \citep{Figer06,RSGC1paper}, Cl~1813-178
\citep{Messineo08}, and Westerlund~2 \citep{Aharonian07}. Finally, in
a follow-up of HESS~J1640-465, a number of hard X-ray sources were
detected in the field of Mc81, with one source being perfectly aligned
with the cluster \citep[][ see right panel of
  \fig{fig:wfim}]{Landi06}. Such emission could be explained by either
a recently-formed neutron star or a colliding wind binary, with both
explanations being indicators of youth \citep[for an analysis of the
  X-ray emission from another young star cluster, Westerlund~1,
  see][]{Clark08}.

Based on this evidence we have extensively followed-up Mc81 with
near-infrared (NIR) photometry and spectroscopy, with a view to
confirming that the object is indeed a young star cluster, and
ultimately to determine the cluster's physical properties.

We begin in Sect.\ \ref{sec:obs} with a description of the
observations, data reduction and analysis steps. In
Sect.\ \ref{sec:res} we describe our results, and show that the object
is indeed a highly extincted star cluster, and estimate its age and
mass. We summarize our results in Sect.\ \ref{sec:summary}.

\section{Observations \& data reduction} \label{sec:obs}

\subsection{Imaging}
Images were obtained with HST/NICMOS on 22 October 2008, as part of
observing programme \#11545 (PI: B.\ Davies). We used the NIC3 camera
which has a field-of-view of 51.2\arcsec$\times$51.2\arcsec\ and a
pixel scale of 0.2\arcsec. We observed the cluster through filters
F160W and F222M, as well as the narrow-band filters F187N and F190N
which are centred on P$\alpha$ and the neighbouring continuum
respectively. In addition to the cluster we observed a nearby control
field through the F160W and F222M filters in order to characterize the
foreground population. The fields of observation are indicated in
Fig.\ \ref{fig:wfim}.

Our observations used a spiral dither pattern with six pointings with
offset distance was set to 5.07\arcsec. This sub-pixel dithering
technique was designed to minimise the impact of non-uniform
intra-pixel sensitivity on our photometry. The MULTIACCUM read modes
were used, with the sampling sequences and total integration times
that are listed in Table \ref{tab:samp}.

Our reduction procedure followed the guidelines of the NICMOS Data
Handbook v7.0. The standard reduction steps of bias subtraction,
dark-current correction and flat-fielding were performed using {\sc
  calnica}. Before mosaicing, each dithered observation was subsampled
onto a 3$\times$ finer grid using bi-linear intepolation to account
for the sub-pixel dithering. 

Photometry was extracted from the images using the {\sc starfinder}
package which run within IDL \citep{STARFINDER}, in conjunction with
point-spread functions (PSFs) which were computed for each filter
using {\sc tinytim}. {\sc starfinder} uses these PSFs to locate stars
within each image, and we employed two iteration cycles to fine-tune
the astrometry and photometry. Since our fields of observation are not
very crowded, we did not use the deblending algorithm, as this was
found to produce many false detections. Uncertainties and completeness
limits were determined by inserting fake stars into each image and
measuring the recovery rate. 

\begin{figure*}
  \centering
  \includegraphics[width=18cm,bb=15 8 672 280,clip]{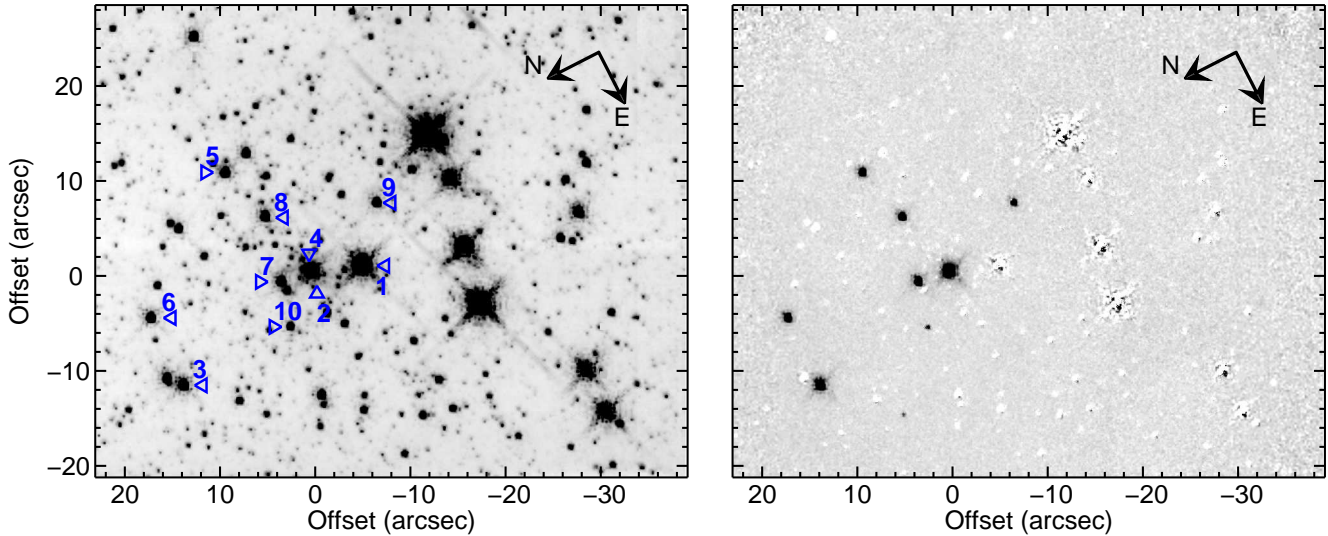}
  \caption{NICMOS images of the cluster. {\it Left}: image of the
    cluster taken through the F222M filter. The two stars for which we
    have spectra, as well as the other emission-line stars, are
    indicated by the blue triangles. {\it Right}: the
    difference image (F187N-F190N), which highlights the emission-line
    stars. The arrows in the top-right of each image indicate the
    orientation.  }
  \label{fig:colfig2}
\end{figure*}

\begin{figure*}
  \centering
  \includegraphics[width=18cm,bb=34 22 436 336,clip]{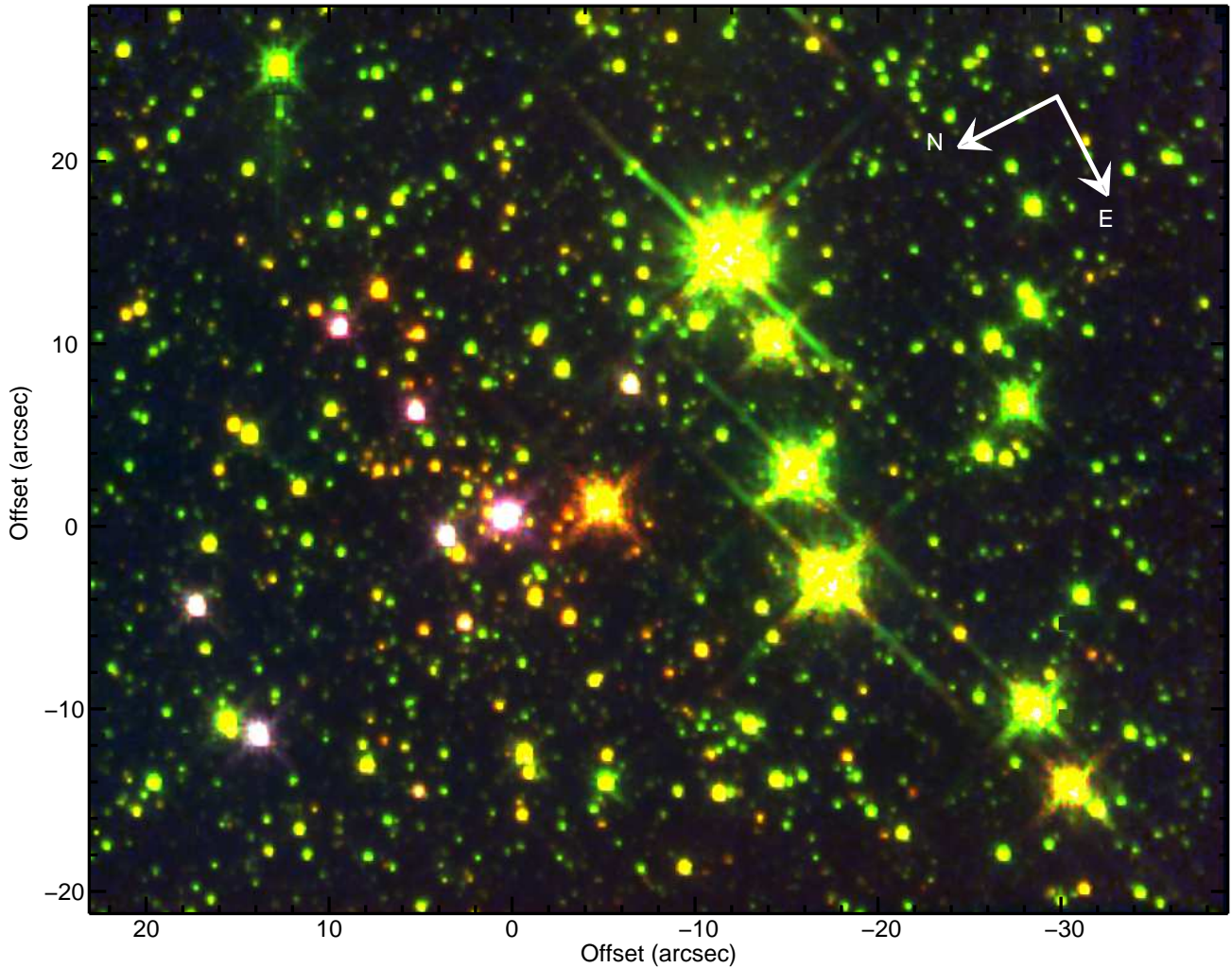}
  \caption{Three-colour image of the cluster from the
    NICMOS data, with colours as follows -- R=F222M, G=F160W,
    B=(F187N-F190N). Therefore, the highly reddened cluster stars
    appear red/orange, and the cluster's emission-line stars appear
    pink/megenta.}
  \label{fig:3col}
\end{figure*}

\begin{table}
  \centering
  \caption{Read-sequences and total integration times employed for each filter
    during the NICMOS observations. }
  \begin{tabular}{lccccc}
    \hline
    \hline
    Filter & SAMP-SEQ & NSAMP & $T_{\rm int}$ (s) \\
    \hline
    F160W  & STEP2    & 15    &  144 \\
    F222M  & STEP8    & 12    &  336 \\
    F187N  & STEP8    & 10    &  240 \\
    F190N  & STEP8    & 10    & 240  \\
    \hline \\
  \end{tabular}
  \label{tab:samp}
\end{table}

\subsection{Spectroscopy}
Spectroscopic data were taken on the nights of 2009-4-11 and 2009-5-4
as part of ESO observing programme 083.D-0765(A) (PI: E.~Puga), using
the ISAAC spectrograph on the VLT \citep{Moorwood98} in `medium'
resolution mode. Our targets were the stars labelled `1' and `3' in
\fig{fig:colfig2}, which were determined to be likely cluster members
based on their photometric properties (see later). We used the
0.8\arcsec\ slit at three different central wavlengths: 1.71\um,
2.09\um, and 2.21\um, providing a spectral resolution of $\Delta
\lambda / \lambda \sim$4,000. The DIT$\times$NDIT$\times$NINT
combination for each wavelength setting was (8$\times$8$\times$30s),
(8$\times$8$\times$32s), (16$\times$8$\times$24s) respectively. The
observations were taken in a ABBA pattern to isolate and subtract sky
emission features. In addition to the target stars we observed the
B9\,{\sc v} stars Hip090248 and Hip091286 as measures of the telluric
absorption, as well as the usual observations of flat-fields, dark
frames and arcs for wavelength calibration. To characterize the
spatial distortion, a bright field star was stepped along the slit and
re-observed multiple times.

The data reduction procedure began with subtraction of nod pairs to
remove sky emission, and division by a normalized flat-field. The 2-D
frames were then rectified onto an orthogonal grid, using the
stepped-star and arc frames to characterize the distortion in the
spatial and dispersion directions respectively. This process also
wavelength calibrates the data, with r.m.s.\ residuals which were found to
be less than a tenth of a resolution element ($\sim$10\kms). After
rectification, the spectra were extracted and combined. 

The telluric standard spectra had their intrinsic \hi\ absoption
removed by fitting the lines with Voigt profiles. The telluric spectra
were then cross-correlated with the target spectra in the region of
isolated telluric features to correct for any residual sub-pixel
shifts, before the target spectra were divided through by the telluric
spectra.


\begin{figure*}
  \centering
  \includegraphics[width=18cm,bb=0 10 730 369,clip]{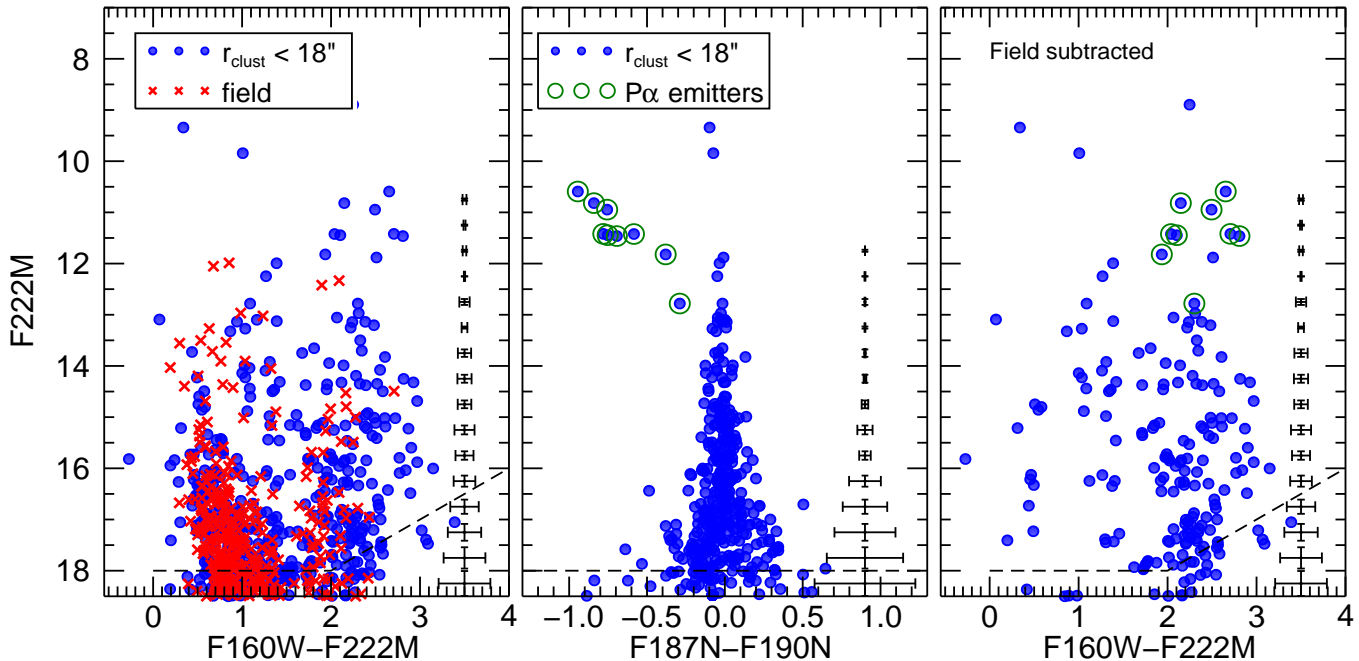}
  \caption{Colour-magnitude diagrams for Mc81 from the NICMOS
    photometry.{\it Left}: The stars within 15\arcsec\ of the cluster
    centre, which we define as the position of Mc81-2, compared to
    stars in a nearby control field of the same angular size. {\it
      Centre}: the P$\alpha$ emission of stars within 15\arcsec\ of
    the cluster centre, as determined from the colour ($m_{187} -
    m_{190}$). Stars with significant emission are marked with green
    circles. {\it Right}: the same as the left panel, after
    the cluster field has been decontaminated of foreground stars
    using the control field observations. The long-dashed lines in
    each figure show the 50\% completeness levels.}
  \label{fig:cmd}
\end{figure*}

\section{Results \& Analysis} \label{sec:res}

\subsection{Photometry} \label{sec:photom}

The NICMOS images of Mc81 can be seen in \fig{fig:colfig2}. The left
panel shows the F222M image of the cluster. In the right panel, we
show the difference image [F187N-F190N], which clearly highlights 9
stars with significant \Pa\ emission\footnote{It is possible that
  \heii\ emission contributes to the flux in this band, especially if
  the stars are WRs.}. This strongly suggests that these are hot stars
with strong winds, and are therefore likely to have ages $\la$10Myr.

In \fig{fig:3col} we show a 3-colour RGB image of F222M (red), F160W
(green), and [F187N-F190N] (blue). Around the coordinate centre, there
is a clear group of highly reddened stars which form the putative
cluster. In this colour scheme, the emission line stars, which are
also heavily reddened, show up as pink/magenta. From these data there
is already strong evidence for a young, highly reddened cluster of
stars in the field of Mc81. The four bright stars to the south of the
cluster have a green/yellow colour, indicating that they are in the
foreground. Ironically, it is likely that these four stars in part
triggered the cluster detection algorithm used by
\citet{Mercer05}. These authors claim that they detected an
association of 65 stars, though the positions of these stars are not
listed, so it is not possible for us to check whether the cluster
detected by Mercer et al.'s algorithm bears any relation to the
cluster we describe here. For the sake of clarity and consistency, we
continue to refer to the cluster studied in this work as Mc81.

The results of the NICMOS photometry are shown in \fig{fig:cmd}. The
left panel shows a colour-magnitude diagram of the stars within
18\arcsec\ of the cluster centre, which we define as the location of
the brightest emission-line star. Also shown on the plot are data from
an area in the control field of identical size. There is an obvious
difference between the two regions, with the `cluster' field showing
an excess of stars at ($m_{160}-m_{222}$)$\simeq$2.3. The centre panel
shows the \Pa\ excess stars. Those stars with ($m_{187}-m_{190}$)
colours $<$0.3 and $m_{222} >$13 are defined as `\Pa\ emitters', and
are indicated on the plot. The locations and photometry of these stars
are listed in Table \ref{tab:emit}.

Finally, the right-hand panel of \fig{fig:cmd} shows the same as the
left, after the cluster field has been decontaminated of field stars
using the data from the control-field. We do this by eliminating stars
from the cluster field which have a control-field star close by in
colour-magnitude space. We set this limiting distance to be the larger
of either the cluster field star's 1-$\sigma$ photometric errors, or
0.15mags in colour and 0.1mags in magnitude. We also set the
restriction that no control-field star can eliminate more than one
cluster field star. After decontamination, the cluster sequence at
($m_{160}-m_{222}$)$\simeq$2.3 becomes clearer, though there is still
some scatter.

\begin{table}
  \centering
  \caption{Astrometry and photometry of Mc81-1, plus the emission-line
    stars. Astrometry is taken from the HST observations, and
    comparisons to 2MASS indicate that it is accurate to
    $\sim$1\arcsec. }
  \begin{tabular}{lccccc}
    \hline \hline
ID  &  RA-DEC (J2000) & $m_{160}$ & $m_{222}$ & $m_{187}$ & $m_{190}$\\
\hline
 1 &  16 40 29.83  -46 23 33.9 &  11.14 &   8.90 &   9.67 &   9.75   \\
 2 &  16 40 29.65  -46 23 29.1 &  13.25 &  10.59 &  10.69 &  11.63   \\
 3 &  16 40 30.08  -46 23 11.4 &  12.97 &  10.82 &  10.75 &  11.60   \\
 4 &  16 40 29.65  -46 23 28.7 &  13.44 &  10.95 &  11.15 &  11.91   \\
 5 &  16 40 28.35  -46 23 25.6 &  14.13 &  11.42 &  11.67 &  12.45   \\
 6 &  16 40 29.32  -46 23 11.6 &  13.46 &  11.42 &  11.56 &  12.14   \\
 7 &  16 40 29.60  -46 23 25.6 &  13.55 &  11.45 &  11.56 &  12.31   \\
 8 &  16 40 28.94  -46 23 27.1 &  14.27 &  11.46 &  11.95 &  12.65   \\
 9 &  16 40 29.32  -46 23 38.2 &  13.76 &  11.82 &  12.16 &  12.54   \\
10 &  16 40 30.05  -46 23 24.3 &  15.08 &  12.78 &  13.41 &  13.70   \\
    \hline \\
  \end{tabular}
  \label{tab:emit}
\end{table}

\begin{figure*}
  \centering
  \includegraphics[width=18cm,clip]{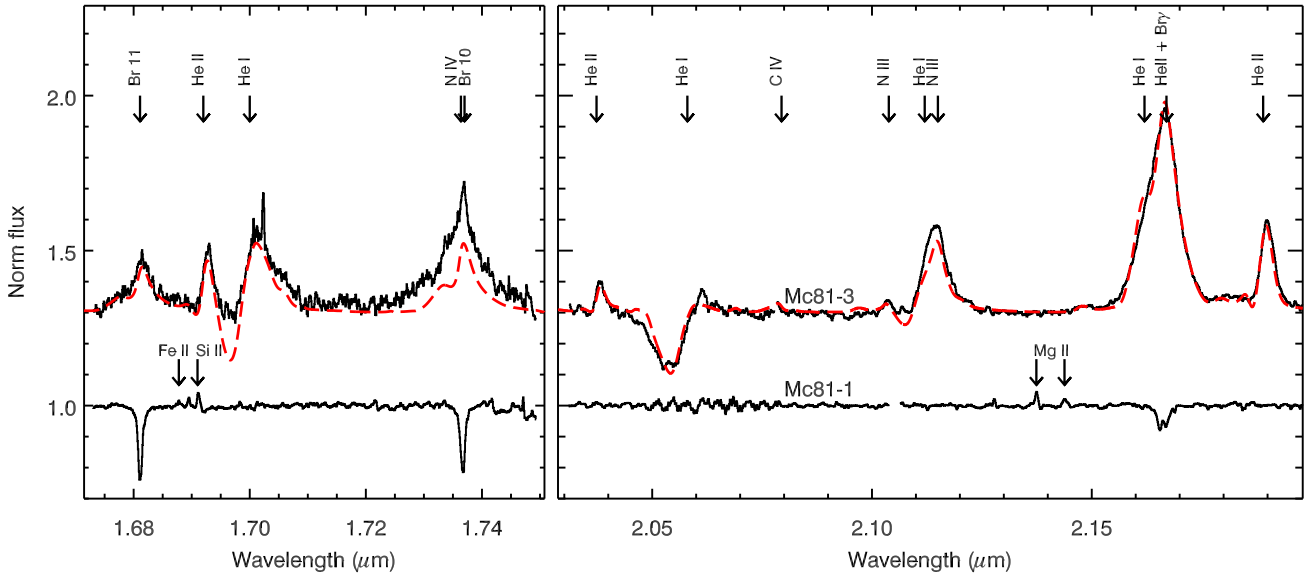}
  \caption{Spectra of the two stars observed. The wiggles seen at
    $\sim$2.05\um\ are due to fringing on the detector, which we were
    unable to remove. }
  \label{fig:spec}
\end{figure*}

\subsection{Spectroscopy}
We now present the results of the spectroscopic observations. To
classify the spectra, we refer to the spectral atlases of
\citet{Hanson05}, \citet{Figer97}, as well as \citet{Crowther-Wd106}.

The spectra of the two brightest stars in Mc81 are shown in
\fig{fig:spec}. The brighter star, Mc81-1, has relatively weak
spectral features. The absorption lines of the Hydrogen Brackett
series are seen, as well as some faint \mgii\ emission. We see no
evidence of He~I in either absorption or emission. This absence of
He~I suggests a spectral type later than $\sim$B5, while the emission
lines of \mgii\ are indicative of a substantial stellar wind, more
typical of supergiants. The `ripples' seen between
2.05-2.08\microns\ are due to poor cancellation of the CO$_2$
absorption. For now, we assign a loose spectral type to this star of
late-B/early-A supergiant. Taking the distance and extinction derived
in Section \ref{sec:extinct}, as well as the bolometric corrections
tabulated by \citet{Blum00} for spectral types B7-A2, we estimate the
luminosity of the star to be in the range $\log(L/L_{\odot}) =
$5.4--5.8, placing the star close to the empirical stellar luminosity
limit at $\log(L/L_{\odot}) \simeq$5.9 \citep{H-D79}. Such stars are
also seen in Westerlund~1 \citep{Clark05}. The star's proximity to the
Humphreys-Davidson limit suggests that the star may be in an unstable
phase of evolution, such as a Luminous Blue Variable or Yellow
Hypergiant phase, though further spectroscopic and photometric
monitoring would be required to verify this.

In contrast to Mc81-1, Mc81-3 has a spectrum rich in strong, broad
emission lines. These lines can be attributed to \hi, \hei, \heii\ and
\niii. The ratio of the \heii\ 2.189\um\ to the complex at 2.115\um,
as well as the absorption of \hei\ 2.189\um, allow us to tightly
constrain the spectral type of this star to be WN7-8.

\subsection{Extinction and distance} \label{sec:extinct}
To calculate the extinction, we first determine the reddening of the
cluster sequence from the right-hand panel of \fig{fig:cmd}. The
average colour of the stars in the decontaminated cluster field is
($m_{160}-m_{222}$)=2.3$\pm$0.3. If we make the approximation that all
main-sequence stars that we detect should have colours of
approximately zero, the observed average colour is due to
extinction. We can then determine the extinction from the following
relation,

\begin{equation}
  A_{\lambda_{2}} = \frac{ E_{\lambda_{1}-\lambda_{2}} }
  { (\lambda_{1}/\lambda_{2})^{\alpha} - 1 }
  \label{equ:extinct}
\end{equation}

\noindent with $\lambda_{1} = $1.60\um\ and $\lambda_{2} = $2.22\um,
i.e. the wavelengths of the NICMOS F160W and F222M filters. The
parameter $\alpha$ has been studied by numerous authors in recent
years \citep[see e.g.][ and references therein]{Stead-Hoare09}, with
the most contemporary measurements converging on $\alpha = -2.0 \pm
0.1$. This therefore implies that the extinction towards Mc81 is
$A_{2.22} = 2.5 \pm 0.5$. Extrapolating this extinction to the optical
is known to be highly uncertain, but we estimate $A_{V} = 45 \pm
15$\footnote{Using the value of $\alpha = -1.53$ from \citet{R-L85},
  we find $A_{2.22} = 3.5 \pm 0.5$, and $A_{V} = 30 \pm 4$}. Mc81 is
therefore one of the most heavily reddened clusters known, with an
extinction comparable to that of the Galactic Centre.

We estimate the distance to the cluster from the radial velocity of
the surrounding molecular cloud. \citet{Caswell-Haynes87} studied the
radio recombination line emission of the two clouds of ionized gas
either side of the cluster, G338.4+0.2 and G338.4+0.1 (see
\fig{fig:wfim}), finding velocities relative to the local standard of
rest of \vlsr=-29\kms\ and -37\kms\ respectively. In addition, a
number of massive young stellar objects (YSOs) and compact
\hii-regions have been detected in the region by the {\it Red MSX
  Source (RMS) Survey} \citep{Hoare05,Urquhart07a,Urquhart07b}, with
an average radial velocity of \vlsr\ = -35.1$\pm$2.8 \kms.

Comparing this average \vlsr\ to the Galactic rotation curve of
\citet{B-B93}, using a Galacto-centric distance of 7.6$\pm$0.3\,kpc
and a rotational velocity of 214$\pm$7\,\kms\ \citep[][ and references
  therein]{K-D07}, we find near and far kinematic distances of 3.8kpc
and 11kpc. To resolve the near/far ambiguity, we refer to the {\it
  Southern Galactic Plane Survey (SGPS)} of H{\sc i}. These data were
recently analysed by \citet{Lemiere09}, who found that for the two
large \hii-regions G338.4+0.2 and G338.4+0.1, H{\sc i} absorption
components could be seen at several velocities up to the tangent point
velocity of 130\kms. This indicates the \hii-regions, and by
association the YSOs and the central star cluster, lay {\it beyond}
the tangent point. From this we conclude that the clusters lay at the
far-side distance of 11\,kpc. At this distance, the uncertainty is
dominated by the systematic uncertainties in the Galactic rotation
curve, which by its nature is difficult to quantify. However, if we
assume that the system may have a peculiar velocity of up to
$\pm$20\,\kms\ \citep{Russeil03}, this gives an uncertainty on this
distance of $\pm$2\,kpc.

We can check this distance by calculating the absolute brightness of
the WNL star and comparing to similar objects. Using the extinction
calculated in the previous section and a distance of 11$\pm$2\,kpc, we
find an absolute magnitude for Mc81-2 of $M_{222} = -7.1 \pm 0.6$. By
comparison, Galactic WNL stars are typically found to have $M_{K} =
5.9 \pm 1.0$ \citep{Crowther-Wd106}. Mc81-2 is therefore somewhat
luminous for its spectral type, though it is within the errors for
other Galactic WNL stars.

\begin{figure}
  \centering
  \includegraphics[bb=10 0 566 453,width=8.5cm]{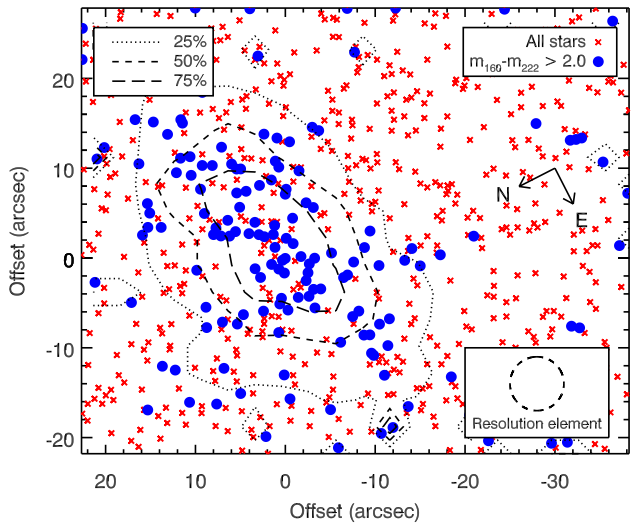}
  \caption{An illustration of the spatial extent of the cluster. Red
    crosses denote all stars in the NICMOS field brighter than the
    50\% detection limit; while blue circles indicate stars with
    colours of {$m_{160} - m_{222} > 2.0$}. The contours indicate
    where the stellar density is 25\%, 50\% and 75\% of the
    maximum. As the data were smoothed to make the contours, the
    circle in the bottom left shows the size of a resolution element. }
  \label{fig:rad}
\end{figure}

\subsection{Cluster size}
In \fig{fig:rad}, we illustrate the physical extent of the
cluster. The figure shows the locations of all stars in the NICMOS
field-of-view which are brighter than the 50\% completeness limit
($m_{222} < 17$), overlayed with those stars which have colours
consistent with the cluster (i.e.\ $m_{160} - m_{222} > 2.0$). The
figure shows that there is a clear overdensity of stars at the
coordinate centre (defined as the position of star Mc81-2). 

To measure the size and morphology of this overdensity, we first made
a map of the stellar density by dividing the field up into square bins
of size 3\arcsec$\times$3\arcsec\ and counting the number of stars per
bin. To reduce noise, this map was then smoothed, such that the
effective resolution of the map was 6\arcsec\ (illustrated in the
bottom corner of \fig{fig:rad}). This resolution size was chosen as a
trade-off between spatial resolution and signal strength, though our
results were robust to changes in this parameter. The ambient stellar
density was found by computing the background level of this map using
the GSFC IDL routine {\sc sky}. Finally, we computed isodensity
contours in this map at percentiles of the maximum stellar density.

We defined the size of the cluster where the stellar density drops to
50\% of its maximum value, which we deem to be roughly equivalant to
its half-light radius\footnote{Ideally, to measure the half-light
  radius one would measure the cumulative surface brightness out to a
  distance where it becomes asymptotic. However, the field-of-view of
  our observations is too small to do this.}. Once deconvolved with
the effective spatial resolution, we find that the cluster has major
and minor axes of 29\arcsec~$\times$~18\arcsec. At a distance of
11kpc, this corresponds to 1.5$\times$1.0\,pc, and so is comparable to
other young Galactic clusters which typically have diameters between
1-2pc \citep[e.g. Trumpler~14, Westerlund~1;][]{Figer08}.



\begin{table}
  \label{tab:cmfgen}
  \centering
  \caption{Best fitting model atmosphere parameters for Mc81-2.}
  \begin{tabular}{lc}
    \hline
    \hline
    Parameter & Value \\
    \hline
    $T_{\rm eff}$ (K)           &  36000 $\pm$ 1000       \\
    $T_{\tau = 20}$ (K)   &  38000 $\pm$ 000         \\
    $v_{\infty}$ (\kms)   &  1350 $\pm$ 100 \\
    $\beta$              & 1.25 \\
    A(H/He)                 & 0.75 $\pm$ 0.25 \\
    $\log (L /$\lsun)           &   6.3 $\pm$ 0.4   \\
    $\log(\dot{M}/$\msunyr)   &  -4.2 $\pm$0.2 \smallskip\\
     \hline \\
  \end{tabular}
\end{table}

\begin{figure}
  \centering
  \includegraphics[width=8.5cm,bb=20 20 600 470,clip]{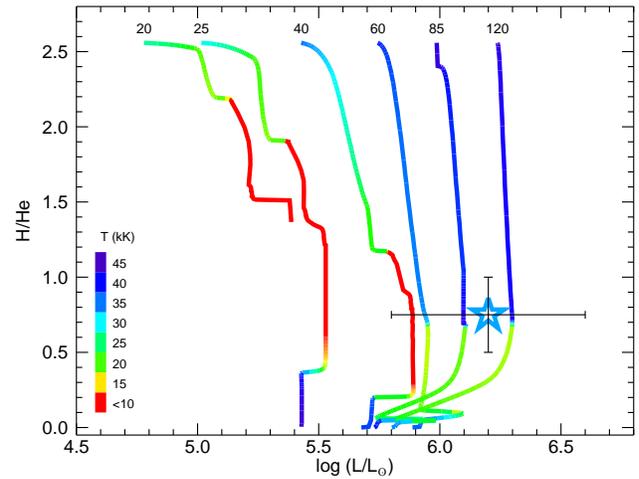}
  \caption{Geneva evolutionary tracks from \citet{Mey-Mae00}. The $x$
    and $y$ axes plot the H/He mass fraction and luminosity
    respectively, while the colour of the track indicates the
    effective temperature at that point in the evolution. The initial
    mass of each track is indicated at the top. Our derived parameters
    for Mc81-2 are indicated by the position and colour of the star
    symbol. Though the symbol intercepts the 120\msun\ track in
    (H/He)-$L$ space, the temperature of the star means that the best
    fit is actually with the 60\msun\ track (see also
    Fig.\ \ref{fig:prob}).}
  \label{fig:logl-hhe}
\end{figure}

\begin{figure}
  \centering
  \includegraphics[width=8.5cm,bb=10 5 605 475,clip]{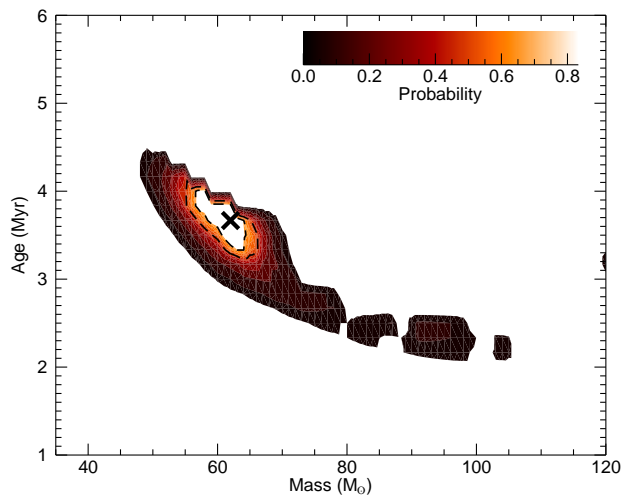}
  \caption{Probability map derived from fitting stellar mass tracks to
  the luminosity, temperature and H/He of the WNL star Mc81-2. The age
  and initial mass of the best fitting model are indicated by the
  cross, and iso-probability contours are drawn at 67\% and 50\%. }
  \label{fig:prob}
\end{figure}

\subsection{Quantitative spectral analysis} \label{sec:cmfgen}

To model the WNL star in Mercer81 and estimate its physical
parameters, we proceed as in \citet{Najarro04,Najarro09}.  Briefly,
we have used CMFGEN, the iterative, non-LTE line blanketing method
presented by \citet{H-M98} which solves the radiative transfer
equation in the co-moving frame and in spherical geometry for the
expanding atmospheres of early-type stars. The model is prescribed by
the stellar radius, \Rstar, the stellar luminosity, \Lstar, the
mass-loss rate, \Mdot, the velocity field, $v(r)$ (defined by the
terminal wind speed \Vinf\ and the wind acceleration parameter
$\beta$), the volume filling factor characterizing the clumping of the
stellar wind, {\it f(r)}, and elemental
abundances. \citet{H-M98,H-M99} present a detailed discussion of
the code.  For the present analysis, we have assumed the atmosphere to
be composed of H, He, C, N, O, Si, S, Fe and Ni. Observational
constraints are provided by the H, K-band spectra of the stars and the
dereddened F166W, F190N and F222M magnitudes.

Given the extreme sensitivity of the $H$ and $K$-Band \hei\ and
\heii\ line profiles ratios in this parameter domain to changes in
temperature, we estimate our errors in the temperature to be below
1000~K. Likewise, the relative strengths between the H and He lines
constrain the H/He ratio to be within 0.5 and 1.0 by mass. The error
on \Rstar\, and hence on \Lstar\ and \Mdot\ is dominated by those in
the assumed distance and the slope of the extinction law.

The best-fitting model is overplotted in \fig{fig:spec}. The model
provides a good fit to the features of the observed spectrum, with the
exception of the Br10 line, which is blended with emission from
\niv/\civ/\oiv. This discrepancy is due primarily to the deficiencies
in the CNO {\sc iv} model atoms, the correction of which is beyond the
scope of the current work. The model's physical parameters are given
in Table \ref{tab:cmfgen}, and are typical for a late-type WN
star. For completeness, we list the temperature at an optical depth of
$\tau = 2/3$, which is comparable to the star's effective temperature;
and the temperature at $\tau = 20$ which is comparable to the
hydrostatic temperature of stellar structure models. In the following
Section we use these results to estimate the age of Mc81-3, and
therefore of the cluster itself.

\subsection{Cluster age}
In order to contrain the age of the cluster, we make a quantitative
comparison between the physical properties of the WNL star Mc81-3 and
the predictions of stellar evolutionary models. Under the assumption
that the star and the host cluster are coeval, we can then estimate
the cluster's age.

For this analysis, the models we have chosen are those of
\citet{Mey-Mae00} which are optimized for massive stars. In our
method, we linearly interpolate these mass tracks at intervals of
1\msun\ and $10^5$yrs. For each point on each interpolated mass track,
we then calculate the probability that there is a match between the
mass track and Mc81-3, based on the star's luminosity $L_{\star}$,
temperature $T$, and H/He ratio $A(\rm He)$ derived in the previous
Section. For the star's temperature, we use the temperature at an
optical depth of $\tau=20$, since this is more comparable to the
hydrostatic temperature calculated by \citet{Mey-Mae00}.

\Fig{fig:logl-hhe} shows the inter-related behaviour of the three
variables \lstar, $T$ and H/He for models with a range of initial
stellar masses. We also plot the derived physical parameters of
Mc81-3. The plot shows that, although the star's luminosity and H/He
ratio place it on the 120\msun\ mass track, the temperature of the
star (illustrated by the colour of the plotting symbol) more closely
matches the 60\msun\ track.

We assume that the errors on $L_{\star}$, $T$ and $A(\rm He)$ are
gaussian, and therefore the probability $p$ of a match between Mc81-3
and the mass track of initial mass $m$ at time $t$ is given by,

\begin{equation}
  p(m,t) = \prod_{i} \exp\displaystyle\left(- \frac{(\mathcal{M}_{i} -
    \mathcal{O}_{i})^2}{2\sigma_{\mathcal{O}_{i}}^{2}} \right)
  \label{equ:prob}
\end{equation}

\noindent where $\mathcal{O}$ is the observed quantity (either
$L_{\star}$, $T$ or $A(\rm He)$), $\sigma_{\mathcal{O}}$ is its
associated uncertainty, and $\mathcal{M}$ is the corresponding
quantity predicted by the model mass track. Each term is therefore
weighted by its associated uncertainty. 

In \fig{fig:prob} we plot how the probability varies across the 2-D
plane of mass and stellar age for the rotating models of
\citet{Mey-Mae00}. The maximum probability ($p=0.75$) is obtained for
an initial mass of $M_{\star}=$62\msun\ and an age of 3.7Myr. The
morphology of the iso-probability contours are highly non-gaussian, so
for the experimental uncertainty we cannot simply compute the standard
deviation. Instead, we take the iso-probability contour at 50\%
and determine the minimum and maximum values of mass and age for that
contour. In this way, for Mc81-3 we find $M_{\star} = 62 ^{+6}_{-7}$\msun\
and an age of $3.7^{+0.4}_{-0.5}$Myr. Using the non-rotating versions
of the same stellar evolution models produces a slightly different
morphology to the probability distribution, with a reduced
probability, but with a best-fitting age and mass that do not differ
significantly from that derived using the rotating models. 

The value we obtain for Mc81-3's age can be understood through a
simple qualitative analysis of the star's parameters. The high
luminosity clearly favours high initial masses, and therefore a young
age. In addition, the He enrichment indicates an object which is in an
advanced evolutionary state, and so older than $\sim$2Myr, but younger
than the total lifetime of a high-mass star, and so therefore younger
than $\sim$5Myr.


\subsubsection{The impact of binary evolution on our derived cluster age}

Throughout this analysis we have assumed that Mc81-3 has evolved as a
single star. However, there is the possibility that the star is in an
interacting binary: both \citet{Landi06} and \citet{Funk07} have
detected X-ray emission from the centre of the cluster, which could be
evidence of a colliding wind binary system. As shown by
\citet{Eldridge09} for the case of $\gamma^2$~Vel, including the
effects of binary evolution in the analysis can alter the derived age
of a star.

In the case of Mc81-3, the star's high luminosity places a strong
constraint on the initial mass of the star, and hence on the upper
limit of its age. Though binary evolution can affect the surface
abundances and temperature of a star, and prolong its lifetime, it is
unlikely to increase the star's maximum luminosity by more than
$\sim$0.1dex, which is governed primarily by the initial stellar mass
(all other parameters being equal) \citep{Eldridge08}. An exception to
this would be if two stars merged to produce a completely rejeuvenated
and more massive star. In the absence of any evidence for such an
event in the history of Mc81-3, we maintain that the upper limit to
the star's age is that derived in the previous paragraph.

The lower limit to the cluster age may be reduced if Mc81-3 is in an
interacting binary system. Mass transfer from the primary to the
secondary star may speed up the rate at which H is depleted from the
primary's surface. In this case, we would underestimate the stellar
(and hence cluster) age by using single star evolution models in our
analysis. 

However, a simple morphological analysis of the nebula surrounding
Mc81, and a comparison to similar systems, serves as a sanity check on
our age estimate. The cluster is located at the centre of a cavity,
which was presumably evacuated by the winds, ionizing radiation and
SNe explosions of the most massive stars in the cluster. At the
periphery of the cavity evidence of further generations of star
formation is seen, which may or may not have been triggered by
feedback from the cluster. This morphology is reminiscent of other
cluster + nebula systems such as G305, NGC~3603 and NGC~346 to name
but a few. Ages of these other systems are commonly found to be 2-4Myr
\citep{Danks-paper,Harayama08,Bouret03}, and therefore are consistent
with our estimate for Mc81.



\subsection{Cluster mass}

For the cluster mass, it is difficult to make an accurate estimate
without further spectroscopy of the stars in the cluster. We can
however make a rough estimate of the cluster mass from the emission
line stars. If we assume that {\it all} the nine strong \Pa\ emitters
listed in Table \ref{tab:emit} are WRs, then since the age we derive
for Mc81 is roughly the same as that of Westerlund~1 (Wd1)
\citep[3-5Myr, ][]{Crowther-Wd106,Brandner08} which has 27 WRs, this
suggests that Mc81 may be a factor of $\sim$3 less massive than
Wd1. As most estimates of Wd1's mass are around
$10^5$\msun\ \citep{Clark05,Brandner08} this implies that the mass of
Mc81 is a few $\times 10^4$\msun. We stress however that this is only
an order-of-magnitude estimate; a more precise measurement of the
cluster mass awaits further analysis of its stellar population.

\section{Discussion}

\begin{figure}
  \centering
  \includegraphics[width=8.5cm,bb=0 0 651 651]{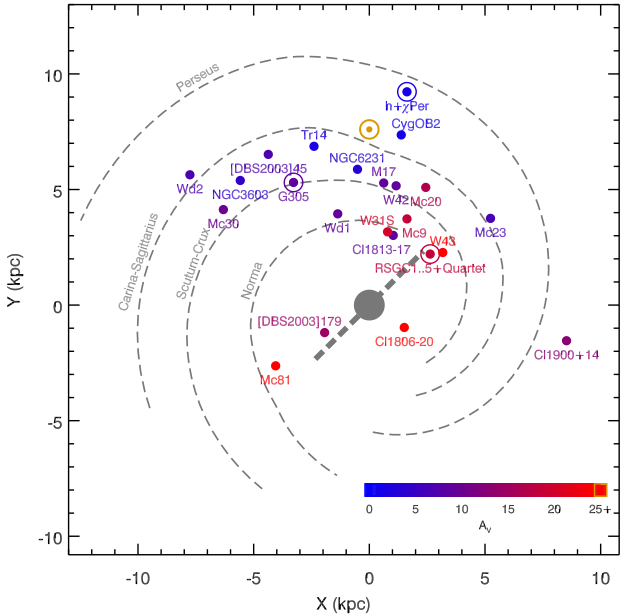}
  \caption{Top down view of the Galaxy, showing the locations of young
    star clusters. The colour of the plotting symbols indicate the
    extinction of each cluster, derived from their near-infrared
    colours and assuming an extinction law slope of -2.0. The spiral
    arms are those defined by \citet{C-L02}. Sites of multiple star
    clusters are indicated by a filled cicle surrounded by an open
    circle. }
  \label{fig:gal-arms}
\end{figure}

\begin{table}
  \label{tab:clust}
  \caption{Young star clusters in the Galactic Plane. The visual
    extinction $A_V$ has been determined homogeneously from each
    cluster's infrared colour excess $E(H-K)$, using the relation
    $A_{V} \simeq 19 \times E(H-K)$, which follows from a NIR
    extinction law with slope $\alpha = -2.0$. See the referenced
    papers for detailed error analysis on the distances. }
  \begin{tabular}{lcccc}
    \hline
    Cluster     &  $l$ (\degr) & $D$ (kpc) &  $A_{V}$ & Ref. \\
    \hline \hline
    Cl1806-20        &  10.0 &  8.7 & 30.3 &   1\\
    W31S             &  10.1 &  4.5 & 20.9 &   2\\
    Cl1813-17        &  12.7 &  4.7 &  9.5 &   3\\
    M17              &  15.0 &  2.4 &  9.5 &   4\\
    Mc9              &  22.8 &  4.2 & 19.0 &   5\\
    W42              &  25.4 &  2.7 &  9.5 &   6\\
    RSGC1..5, Quartet &  26.0 &  6.0 & 19.0 &  7,8,9,10,11,12\\
    W43              &  30.8 &  6.2 & 39.8 &   13\\
    Cl1900+14        &  43.0 & 12.5 & 13.3 &   14\\
    Mc20             &  44.2 &  3.5 & 17.1 &   12\\
    Mc23             &  53.7 &  6.5 &  6.6 &   15\\
    CygOB2           &  80.2 &  1.4 &  1.9 &   16\\
    h+$\chi$ Per     & 135.0 &  2.3 &  1.9 &   17\\
    $[$DBS2003$]$45  & 283.9 &  4.5 &  7.6 &   18\\
    Westerlund 2     & 284.2 &  8.0 &  7.6 &   19\\
    Trumpler 14      & 287.0 &  2.5 &  2.5 &   20\\
    NGC3603          & 291.6 &  6.0 &  4.7 &   21\\
    Mc30             & 298.8 &  7.2 &  10.5 &  22\\
    Danks 1 \& 2     & 305.0 &  4.0 &  9.5 &   23\\
    Mc81             & 338.4 & 11.0 & 41.7 &   This work.\\
    Westerlund 1     & 339.5 &  3.9 &  9.5 &   24\\
    NGC6231          & 343.5 &  1.8 &  3.8 &   25\\
    $[$DBS2003$]$179 & 347.6 &  9.0 & 15.2 &   26, This work.\\
    \hline \\
  \end{tabular}
References: 1:~\citet{Bibby08}; 2:~\citet{Blum01};
3:~\citet{Messineo08}; 4:~\citet{Hanson96}; 5:~\citet{Messineo10};
6:~\citet{Blum00}; 7:~\citet{RSGC1paper}; 8:~\citet{RSGC2paper};
9:~\citet{RSGC3paper}; 10:~\citet{RSGC4paper}; 11:\citet{RSGC5paper};
12:~\citet{Messineo09}; 13:~\citet{Blum99}; 14:~\citet{SGR1900paper};
15:~\citet{Hanson10}; 16:~\citet{Hanson03}; 17:~\citet{Currie10};
18:~\citet{Zhu09}; 19:~\citet{Rauw07}; 20:~\citet{Ascenso07};
21:~\citet{Harayama08}; 22:~\citet{Kurtev07}; 
23:~\citet{Danks-paper}; 24:~\citet{K-D07};
25:~\citet{Raboud97}; 26:~\citet{Borissova08}. \\
\end{table}

\subsection{Location in the Galaxy}
With the many recent discoveries of young star clusters in the Galaxy,
we can now begin to build up a picture of the Galaxy's recent cluster
formation. In \fig{fig:gal-arms} we plot the locations of all known
young Galactic clusters with distances from the Sun greater than
$\sim$2\,kpc. All clusters in the plot are thought to have masses in
excess of $10^3$\msun\ and ages $\la$20Myr. The references for each
data-point are listed in Table \ref{tab:clust}. We have colour-coded
each data-point according to its visual extinction, which we have
calculated in a homogeneous way from each cluster's $E(H-K)$, measured
either from the references listed in \ref{tab:clust} or from 2MASS
photometry, and an extinction law slope of $\alpha = -2.0$ (see
Sect.\ \ref{sec:extinct}).

\Fig{fig:gal-arms} shows that there are now a significant number of
reddened star clusters known at Galactic longitudes between 10$\ga l
\ga$50. This line-of-sight corresponds to the tangent of the
Scutum-Crux arm, as well as the near end of the Galactic Bar
\citep{Benjamin05}. Since one would expect the star formation rate to
be comparatively high in this location of the Galaxy, it is also
reasonable to expect it to be rich in young star clusters. Indeed, the
region hosts five known YMCs, plus a substantial field population of
Red Supergiants, indicating a starburst episode of
$\sim10^6$\msun\ around 20Myr ago
\citep{Garzon97,Lopez-Corredoira99,Figer06,RSGC2paper,RSGC3paper,RSGC4paper,RSGC5paper}.

However, less is known about the opposite side of the Galactic Centre
and far end of the Bar. There are two possible reasons for this:
firstly, the larger distance, high extinction and larger number of
foreground stars (due to the intervening Galactic Bulge) make it more
difficult to pick out clusters in by-eye searches. Indeed, it is
unlikely that Mc81 would have been found were it not for the four
bright foreground stars (see Sect.\ \ref{sec:photom}). Secondly, the
fact that no star clusters are known in this direction means that
investigators are less likely to search this region. This is in
contrast to the near end of the Bar, where the initial discovery of
the cluster RSGC1 in this region by \citet{Figer06} led
to the subsequent discoveries of a further 4 clusters within the same
complex
\footnote{Though
RSGC1 and RSGC2 were first recognized as associations of stars by
\citet{Bica03} and \citet{Stephenson90} respectively, the nature of
each cluster was not understood until later.}.

From our distance estimate of Mc81, we can place this cluster close to
where we suppose the far-end of the Bar may be, assuming an azimuthal
angle of 44\degr\ and a bar length of 4.4\,kpc
\citep{Benjamin05}. Another cluster nearby, which may too trace the
end of the Bar, is [DBS2003]179. The distance to this cluster is not
well known, and is based on spectro-photometric distance estimates for
stars with unknown luminosity classes \citep{Borissova08}. We have
reassesed the distance to this cluster using a similar methodology
that we have presented here for Mc81. Specifically, we assume that the
cluster is physically associated to its nearby molecular cloud, and
use the massive YSOs detected in the cloud to determine the systemic
radial velocity. We then use the SGPS survey
\citep{McClure-Griffiths05} to measure the velocity spectrum of the
intervening \hi\ gas to resolve the distance ambiguity. The average
radial velocities of the YSOs (\vlsr=-36$\pm$3\kms), combined with the
\hi\ absorption which is seen up to tangent-point velocities of
130\kms, give a far-side kinematic distance of 9\,kpc, again with a
$\pm$2\,\kms\ uncertainty to allow for deviations from the Galactic
rotation curve (see Sect.\ \ref{sec:extinct}). This is within the
errors of the spectro-photometric distance of 7.9\,kpc derived by
\citet{Borissova08}.

These two clusters -- Mc81 and [DBS2003]179 -- are then the first
young star clusters to be discovered in this region of the Galaxy. In
addition to these clusters a group of giant \hii-regions, of which
G338.4+0.1 is one \citep{Russeil03}, suggest that this region may be
an active star-formation site, similar to the region of the
Scutum-Crux tangent at the opposite end of the Bar. As such, future
targeted surveys of this region may unearth a number of other such
objects.


%
%

\subsection{Association with HESS~1640-465}

As was noted in the introduction, the Mc81 cluster is located only a
few arcminutes from the TeV source HESS~1640-465, which is likely to
be powered by a neutron star. If the two were associated, it would
allow us to estimate the initial mass of the neutron star's
progenitor. Here we discuss the possible association between the two
objects.

HESS~1640-465 is known to have a counterpart source at GeV energies
\citep{Slane10}, and in the X-ray, detected by {\it Swift}, {\it XMM}
and {\it Chandra} \citep{Landi06,Funk07,Lemiere09}. Analysis of the
{\it XMM} X-ray spectrum yielded a column density of $n_{\rm H} =
6.1^{+2.1}_{-0.6} \times 10^{22}$cm$^{-2}$ or $3.6^{+1.1}_{-0.8}
\times 10^{22}$cm$^{-2}$, depending on whether a power-law or absorbed
black-body model was used \citep{Funk07}. However, \citet{Lemiere09}
analysed the {\it Chandra} data and found that they required a higher
column density of $1.4 \times 10^{23}$cm$^{-2}$ to fit the
data. Assuming a standard calibration between $n_{\rm H}$ and optical
extinction $A_{V}$ of $n_{\rm H} = 1.8 \times 10^{21}
A_{V}$\,cm$^{-2}$ \citep{Predehl-Schmitt95}, this implies an visual
extinction of between 20 and 70 mags, depending on the model for the
X-ray emission. Though the errors are large, this is consistent with
our measurment of the extinction to the cluster of $A_{V} = 45 \pm
15$.

From this evidence, it seems likely that HESS~1640-465 is associated
with the G338.4+0.1 \hii-region surrounding Mc81. However, its
connection with the cluster itself is not clear. If the central source
of HESS~1640-465 is a neutron star (as seems likely), and the
progenitor {\it was} born with the cluster but was ejected, there are
two possibilities: either the progenitor was dynamically ejected from
the cluster; or it received a kick from the supernova (SN). The
location of HESS~1640-465 at the centre of SNR~338.3+0.0
\citep{Green04} provides circumstantial evidence against the latter
explanation, since this suggests that the progenitor exploded close to
its present location. If the progenitor formed with the cluster but
was ejected at a time $t_{\rm ej}$ ago, the ejection velocity is
$\simeq 20 (t_{\rm ej}/{\rm Myr})$\kms (assuming a projected distance
of 22\,pc). Therefore, it is entirely plausible that the progenitor
star formed along with the rest of the stars in Mc81 and was
dynamically ejected during the formation of the cluster.

Finally, there is the possibility that HESS~1640-465 formed out of the
same molecular cloud as Mc81, and at a similar time, but that the two
formed independently of one another. The morphology of the G338.4
region suggests inside-out star-formation, with the $\sim$3Myr old
cluster in the centre and a series of YSOs and UC-\hii\ regions at the
periphery of the surrounding cavity, which have ages of a few $\times
10^5$yrs \citep{simgal}. The location of HESS~1640-465 does not fit
this picture, since the progenitor star must have formed at least 2Myr
ago, which is before the first SNe occured in Mc81. However, there are
other known instances of `multi-seeded' star formation, where collapse
occurs at multiple causally-unrelated sites across the host GMC
\citep[e.g.\ W51,][]{Clark09}.

In summary, we conclude that HESS~1640-465 is likely associated with
the star-formation region of G338.4+0.1. However, we are unable to
make a definitive association with the star cluster Mc81, and so we
are unable to use the age of the cluster to estimate the mass of the
neutron star's progenitor, as we were in the cases of e.g.\ RSGC1 and
Cl~1900+14 \citep{RSGC1paper,SGR1900paper}.



%
%

%




%

\section{Summary} \label{sec:summary}

We have provided a near-infrared photometric and spectroscopic
investigation of the candidate star cluster Mercer~81 (Mc81). We find
that that a highly extincted ($A_{V} = 45 \pm 15$) cluster exists in
the field identified by Mercer et al.\ (2005), but that the bright
four stars at the centre of the field are unrelated foreground
objects. The cluster is located at the centre of a cavity in a large
\hii-region in the direction of G338.4+0.1, with evidence of on-going
star formation in the periphery of the cloud. Our analysis of the
cluster has revealed nine stars with strong \Pa\ emission, one of
which we identify spectroscopically as a late-type N-rich Wolf-Rayet
star (WNL), in addition to a luminous early A-type supergiant. Via
detailed modelling of the WNL star's spectrum we estimate an age for
the cluster of $3.7^{+0.4}_{-0.5}$Myr. Under the assumption that the
stars with strong line emission are WRs, we have made an
order-of-magnitude estimate of the cluster's mass of $\ga 10^4$\msun.

From a kinematic analysis of the host cloud, we obtain a distance to
the host star-forming complex of 11$\pm$2\,kpc. Our distance estimate
therefore places the G338.4+0.1 complex in the same region of the
Galaxy as the far end of the Galactic Bar. The recent detection of
another star cluster close to this location, as well as other giant
\hii-regions, suggest that this region of the Galaxy may be as active
in star-formation as the opposite end of the Bar, where a $\sim
10^6$\msun\ starburst event is known to have occurred in the last
$\sim$20Myr. A targeted search of the far end of the Bar will likely
uncover many more young star clusters, though the high extinction,
large distance and dense stellar field of the intervening Galactic
Bulge will mean that such a search will require considerable
observational effort.

\section*{Acknowledgments}
We thank the anonymous referee for comments and suggestions which
helped us improve the paper. BD is supported by a fellowship from the
Royal Astronomical Society. This work is in part based on observations
made with the NASA/ESA Hubble Space Telescope, obtained at the Space
Telescope Science Institute, which is operated by the Association of
Universities for Research in Astronomy, Inc., under NASA contract NAS
5-26555. These observations are associated with program
\#11545. Support for program \#11545 was provided by NASA through a
grant from the Space Telescope Science Institute, which is operated by
the Association of Universities for Research in Astronomy, Inc., under
NASA contract NAS 5-26555. This work is in part based on observations
collected at the European Organisation for Astronomical Research in
the Southern Hemisphere, Chile, under programme number
083.D-0765(A). Financial support from the Spanish Ministerio de
Ciencia e Innovaci\'on under projects AYA2008-06166-C03-02 and
AYA2010-21697-C05-01 is acknowledged.

\bibliographystyle{/fat/Data/bibtex/aa}
\bibliography{/fat/Data/bibtex/biblio}

\end{document}